\documentclass{rmaa}
\usepackage{graphics}

\title{Visual Spectroscopy of Asteroids at San Pedro Martir}

\author{A.~Manara and S.~Covino\affil{Osservatorio Astronomico di Brera--Milano} 
M.~Di~Martino\affil{Osservatorio Astronomico di Torino}}

\resumen{En este trabajo se presentan los resultados de la espectroscop\'\i a de 
baja resoluci\'on para tres asteroides Nysa y uno de la familia ``near-Earth''.
Los espectros cubren el rango de longitud de onda entre 5000 y 10000 \AA\ y se 
obtuvieron en el telescopio de 2.1m del OAN en San Pedro M\'artir. A\'un con los 
limites importantes debidos al rango espectral abarcado, el an\'alisis revela 
que dos de los asteroides de la familia Nysa, (2007) McCuskey y (3130) Hillary,
son compatibles con un tipo F mientras que el tercero, (3384) Daliya, pertenece 
al tipo S. Para (3908) Nyx, el asteroide de la clase ``Near-Earth'' (tipo 
``Amor''), se sugiere una clasificaci\'on taxon\'omica consistente con un tipo 
V. Por lo tanto, este objeto podr\'\i a ser el resultado de un impacto con el 
asteroide Vesta.}

\abstract{We present low resolution reflectance spectra over the wavelength 
range 5000--10000\AA\  for 4 asteroids (3 belonging to the Nysa family and one 
near--Earth asteroid) obtained at the San Pedro Martir Observatory (Mexico) 
using the 2.1\,m telescope. Though the limited wavelength range covered often 
makes a classification difficult, the analysis of the available data reveals 
that two asteroids of the Nysa family, (2007) McCuskey and (3130) Hillary, are 
probably of F-type and (3384) Daliya of S-type. Near--Earth asteroid (3908) Nyx 
(Amor asteroid) shows a spectrum, within the limits of our signal--to--noise 
ratio (S/N), consistent with a V--type taxonomic classification and may be a 
fragment excavated from Vesta by an impact.}


\shortauthor{Manara, Covino \& Di Martino}

\shorttitle{Visual Spectroscopy of Asteroids at San Pedro Martir}

\fulladdresses{
\item A.~Manara and S.~Covino: Osservatorio Astronomico di Brera--Milano, via 
Brera 28, I--20100, Milano, Italy (manara,covino@brera.mi.astro.it)
\item M.~Di~Martino: Osservatorio Astronomico di Torino, I--10025 Pino Torinese, 
Italy (dimartino@to.astro.it)}

\SetVolume{...} 
\SetFirstPage{1} 
\SetYear{...}
\ReceivedDate{...} 
\AcceptedDate{...}
\SetMSnumber{...}

\begin{document}

\maketitle

\section{Introduction}

The population of Earth--approaching asteroids has long eluded systematic study
due their small sizes and low brightness. Spectroscopic observations can help
to determine the surface mineralogy of Near Earth Asteroids (NEAs). 
Investigations of asteroid compositions can identify potential parent bodies of 
specific meteorites or meteorite types or objects which have experienced similar 
evolutionary histories. Most meteorites are asteroidal fragments ejected from 
their parent bodies as a consequence of impacts, and channeled into chaotic 
dynamical routes, associated with mean motion and secular resonances. The main 
problem is that approximately 73\% of the meteorites that fall on Earth are 
classified as ordinary chondrites (consisting of grains of olivine and pyroxene 
thought to be only modestly altered during the formation process), which cannot 
be matched with the typical observed reflectance spectra of any common asteroid
taxonomic type (Wetherill \& Chapman 1988). The source of these bodies is still
a matter of great debate. Numerical orbital dynamic simulations (Migliorini
et al. 1998) show that many asteroids in the main belt are driven toward 
Mars--crossing orbits by numerous weak mean motion resonances; in addition, half 
of the Mars--crossing asteroids are injected in Earth--crossing orbits in less 
than 20 million years. Gladman et al. (1996) suggest that even ejecta from
Mars may be consistent with the dynamical constraints imposed by the small
Earth--approachers. There has been a persistent problem of finding a source body
for the ordinary chondrites. Due to their dynamically short lifetimes 
(10--100\,Myr), Near--Earth asteroids must be actively replenished (Wetherill
1985, 1988). It has been argued that S(IV)--type asteroids provide the only
plausible source of parent bodies, with (3) Juno, (6) Hebe and (7) Iris being 
the leading candidates (Gaffey et al. 1993, Broglia et al. 1994, Migliorini et
al. 1997a, 1997b). 

The reflectance spectra of 3 Nysa family asteroids have been measured in
order to investigate the mineralogical characterization of this family. These
observations belong to a systematic campaign to study the peculiar Nysa
family (Zappal\`a et al. 1995).
%since it splits into two major subclusters just below the critical distance
%level used to define families in this region of the belt. 
We have some difficulty in assessing whether the Nysa family can
be considered as a unique group or as the result of the merging of two
independent families, because it is known that in the region of the belt
surrounding the family, there is an unusual concentration of F--type asteroids,
some of them are included into the list of nominal Nysa members, while some
others apparently do not belong to the family.

We carried out spectroscopic observations of Near--Earth asteroid (3908) Nyx and
3 Nysa asteroids family (2007) McCuskey, (3130) Hillary, (3384) Daliya.

\section{Observations and data reductions}

Spectroscopic observations were performed at the San Pedro Martir Observatory
(Mexico) using a 2.1\,m telescope equipped with a Boller \& Chivens spectrograph
and a CCD--Tektronix TK--1024 AB detector at the f7.5 focus with a dispersion
of 8\,\AA/pixel in the wavelength direction. The grating used was a 150\,gr/mm 
with a dispersion of 326\,\AA/mm; also a GG\,455 filter was used (blaze 3:26). 
The useful spectral range is about $4800 < \lambda < 10000$.

The slit width (2" and 2".5) has been chosen to minimize the consequences of
atmospheric differential refraction and to reduce the loss of light at both
ends of the spectrum. We observed in 1996 September 3--6, but only the 
first night was good for the observations. 

\begin{table*}[ht]
\caption{Observing Conditions}
\begin{center}
\begin{tabular}{|ccrrcccc|}
\hline \hline
{\bf Number \& Name} & {\bf UT} & {\bf $\alpha_{\mbox{2000}}$} & {\bf 
$\delta_{\mbox{2000}}$} & {\bf R} & {\bf $\Delta$ } & {\bf Phase} & {\bf $m_V$ } 
\\
         & {\em 1996 Sep. }   & {\em h\,m\,sec }           & {\em d\,m\,s} & 
{\em AU} & {\em AU} & &             \\
\hline
(2007)~McCuskey   & 4.39375  & 00\,47\,39               & +03\,42\,01   &  2.641 
&   1.723 &  11.3 &  15.8   \\
%(2744)~Birgitta  & 4.44930   & 01\,03\,23               & +18\,54\,00   &  
%1.538 &   0.628 &  25.6 &  15.9   \\
(3130)~Hillary  & 4.35278    & 22\,58\,28               & -13\,48\,51   &  1.982 
&   0.977 &   3.4 &  14.5 \\
(3384)~Daliya  & 4.30555     & 22\,34\,32               & -10\,53\,06   &  2.085 
&   1.079 &   2.8 &  15.7 \\
(3908)~Nyx  & 4.23333        & 20\,54\,11               & -03\,25\,56   &  1.174 
&   0.186 &  25.2 &  15.2 \\
\hline \hline
\end{tabular}
\end{center}
\label{tab:uno}
\end{table*}

The observational circumstances are listed in Table\,\ref{tab:uno}.

Column\,1 gives the observed object, column\,2 the date of observations, 
column\,3 and 4 RA and DEC of the object, column\,5 and 6 Sun and Earth 
distance, column\,7 phase angle and the last column the visual apparent 
magnitude. 

The spectral data reduction was performed using the ESO--MIDAS  package and 
taking much care to ensure a proper calibration of the spectra. The bias level 
of each night was determined through an average of the many bias images taken at 
night. This averaged bias was then subtracted from each frame and 
pixel--to--pixel variations were removed by dividing the resulting image by a 
normalized medium flat field. The MIDAS ``long'' context was used to sum the 
pixel values within a specified aperture and to subtract the background level. 
Wavelength calibration was performed several times during each night using a 
He--Ar lamp, and spectra were corrected for airmasses by using the mean 
extinction curve of San Pedro Martir (Buzzoni 1994).

This correction was checked by comparing the same analog star taken at different 
air mass and the differences were negligible. Since each analog was observed 
several times we also reduced each asteroid spectrum with the solar analog taken 
at the same air mass (or as near as possible). Again no difference could be 
observed, which confirms the quality of the data. Two solar analog, 16~Cyg\,B 
and 
HD\,191854 (Hardorp 1978) were observed to compute reflectivities, since these 
are solar analogs which closely match the spectra of the Sun. The ratios 
between the spectra of the two solar analogs for the night of September 3th show 
no substantial variation. The influence of different solar analogs on the 
resulting spectra has also been checked, showing negligible differences. The 
obtained reflectance spectra are normalized at 7000\,\AA.

\section{Results}  

\paragraph{Asteroids of Nysa family}

\begin{figure}
\begin{center}
\rotatebox{270}{\includegraphics[height=\columnwidth]{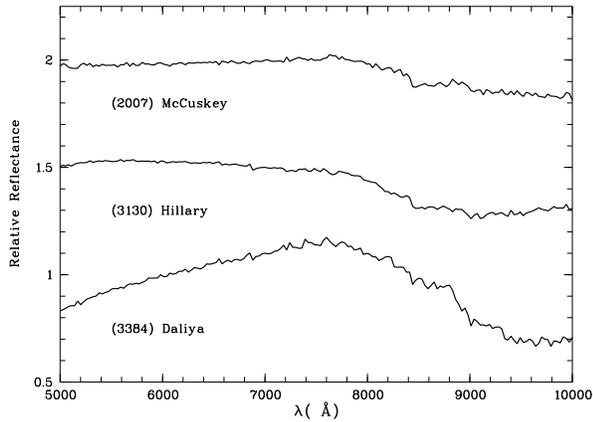}}
\end{center}
\caption{Reflectance spectra curves for asteroids (2007) McCuskey, 
(3130) Hillary and (3384) Daliya.}
\label{fig:1}
\end{figure}

The spectra of (2007) McCuskey and (3130) Hillary are similar, therefore
this confirms their membership on the same group (Fig.\,\ref{fig:1}). These 
spectra are representative of subgroup of objects of the Nysa family belonging 
to the F taxonomic class. This conclusion can be also obtained by comparing
the spectral reflectance curve of these two asteroids with those (in 
particular asteroids 2391, 4026, 3485, 3228, 3064, 1076) shown in  
Cellino et al. (2000) and in Xu et al. 1995.
Concerning asteroid (3384) Daliya, his spectrum (Fig.\,\ref{fig:1}) shows a 
curve similar to S--type objects (Cellino et al. 2000) and therefore is a 
representative of the subgroup of the Nysa family belonging to S taxonomic 
class. In any case it is also possible that (3384) Daliya may be an interloper 
of the Nysa family (Zappal\`a et al. 1995).

\paragraph{Asteroid (3908) Nyx}

\begin{figure}
\begin{center}
\rotatebox{270}{\includegraphics[height=\columnwidth]{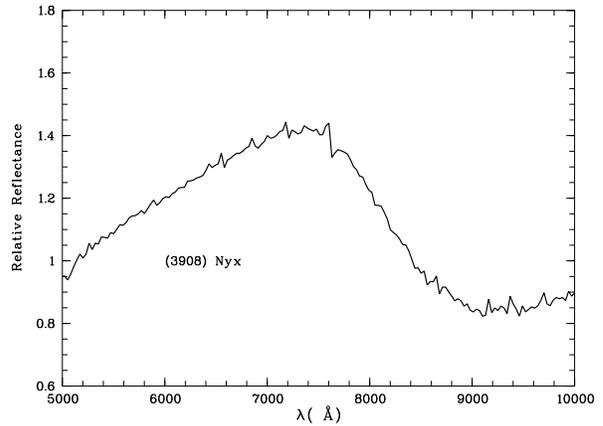}}
\end{center}
\caption{Reflectance spectrum curve for asteroid (3908) Nyx.}
\label{fig:2}
\end{figure}

This near-Earth asteroid, classified as V--Type by Tholen and Barucci (1989), 
has been spectroscopically studied by McFadden et al. (1989) and Luu \& Jewitt 
(1990). Our spectrum (Fig.\,\ref{fig:2}) confirms the strong absorption feature 
centered at about 9500\AA. The similarity between the spectrum of 3908 Nyx and 
those of objects belonging to the Vesta Family (Binzel \& Xu, 1993) suggests 
that also this asteroid is a chip of a Vesta-like parent body (Cruikshank et al. 
1991).

\paragraph{Acknowledgments} We are grateful to the staff at San Pedro Martir 
Observatory (Mexico) for their support during the observing runs.

\end{document}